\documentclass[Proceedings,11pt,letterpaper]{ascelike}
\usepackage[pdftex]{graphicx}
\usepackage{times}
\usepackage{amsmath, amssymb, amstext}
\usepackage[boxed,algonl]{algorithm2e}
\newcommand{\Sa}{\ensuremath{\sigma_a}}
\newcommand{\Shi}{\ensuremath{\hat\sigma_i}}

\newcommand{\Gi}{\ensuremath{g_{0i}}}
\newcommand{\Ge}{\ensuremath{g_{0e}}}
\newcommand{\Ei}{\ensuremath{\dot\epsilon_{0i}}}
\newcommand{\Ee}{\ensuremath{\dot\epsilon_{0e}}}
\newcommand{\En}{\ensuremath{\epsilon_{n}}}
\newcommand{\Pci}{\ensuremath{p_{i}}}
\newcommand{\Pce}{\ensuremath{p_{e}}}
\newcommand{\Qci}{\ensuremath{q_{i}}}
\newcommand{\Qce}{\ensuremath{q_{e}}}

\newcommand{\TIV}{\ensuremath{\theta_{IV}}}
\newcommand{\Ses}{\ensuremath{\hat\sigma_{es0}}}
\newcommand{\Ges}{\ensuremath{g_{0es}}}
\newcommand{\Ees}{\ensuremath{\dot\epsilon_{es0}}}

\newcommand{\Bnabla}{\ensuremath{\boldsymbol{\nabla}}}

\newcommand{\Bsig}{\ensuremath{\boldsymbol{\sigma}}}

\newcommand{\Ba}{\ensuremath{\mathbf{a}}}

\newcommand{\Bf}{\ensuremath{\mathbf{f}}}
\newcommand{\Bg}{\ensuremath{\mathbf{g}}}
\newcommand{\Bm}{\ensuremath{\mathbf{m}}}
\newcommand{\Bn}{\ensuremath{\mathbf{n}}}

\newcommand{\Bv}{\ensuremath{\mathbf{v}}}

\newcommand{\Bx}{\ensuremath{\mathbf{x}}}

\newcommand{\BD}{\ensuremath{\mathbf{D}}}

\newcommand{\BG}{\ensuremath{\mathbf{G}}}

\newcommand{\Tint}{\ensuremath{\text{int}}}
\newcommand{\Text}{\ensuremath{\text{ext}}}

\newcommand{\Half}{\ensuremath{\frac{1}{2}}}

\newcommand{\Grad}[1]{\ensuremath{\Bnabla #1}}

\begin{document}

\setlength{\textwidth}{6in}
\setlength{\oddsidemargin}{.25in}
\thispagestyle{empty}

\begin{figure}
\flushright
\includegraphics[width=3.25in]{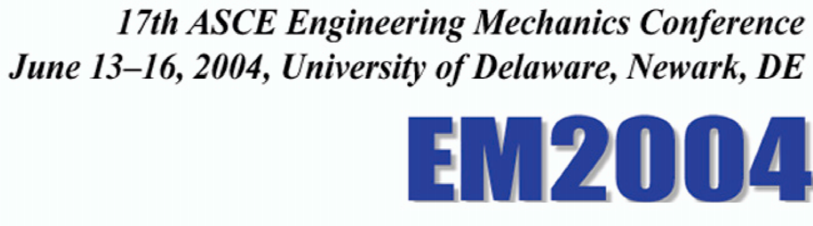}
\vspace*{-1in}
\end{figure}

%
\title{MATERIAL POINT METHOD SIMULATIONS OF FRAGMENTING CYLINDERS}
\author{
Biswajit Banerjee %
\thanks
{Department of Mechanical Engineering, University of Utah, Salt Lake City, UT 84112, USA}%
%
}
\maketitle
\begin{abstract}
  Most research on the simulation of deformation and failure
  of metals has been and continues to be performed using the finite element
  method.  However, the issues of mesh entanglement under large deformation,
  considerable complexity in handling contact, and difficulties encountered
  while solving large deformation fluid-structure interaction problems have
  led to the exploration of alternative approaches.  The material point 
  method uses Lagrangian solid particles embedded in an Eulerian grid.  
  Particles interact via the grid with other particles in the same body, with 
  other solid bodies, and with fluids.  Thus, the three issues mentioned in 
  the context of finite element analysis are circumvented.

  In this paper, we present simulations of cylinders which fragment due to 
  explosively expanding gases generated by reactions in a high energy material
  contained inside. The material point method is the numerical method chosen 
  for these simulations discussed in this paper.  The plastic deformation of 
  metals is simulated using a hypoelastic-plastic stress update with radial 
  return that assumes an additive decomposition of the rate of deformation 
  tensor.  Various plastic strain, plastic strain rate, and temperature 
  dependent flow rules and yield conditions are investigated.  Failure at 
  individual material points is determined using porosity, damage and 
  bifurcation conditions.  Our models are validated using data from 
  high strain rate impact experiments.  It is concluded that the material 
  point method possesses great potential for simulating high strain-rate,
  large deformation fluid-structure interaction problems.
\end{abstract}
%
%
\section{Introduction}
  The goal of this work is to present results from the simulation of
  the deformation and failure of a steel container that expands under 
  the effect of gases produced by an explosively reacting high energy 
  material (PBX 9501) contained inside.
  
  The high energy material reacts at temperatures of 450 K and above,  This
  elevated temperature is achieved through external heating of the steel 
  container.  Experiments conducted at the University of Utah have shown that
  failure of the container can be due to ductile fracture associated with 
  void coalescence and adiabatic shear bands.  If shear bands dominate the
  steel container fragments, otherwise a few large cracks propagate along the
  cylinder and pop it open.

\begin{figure}
    \centering
    \scalebox{0.5}{\includegraphics{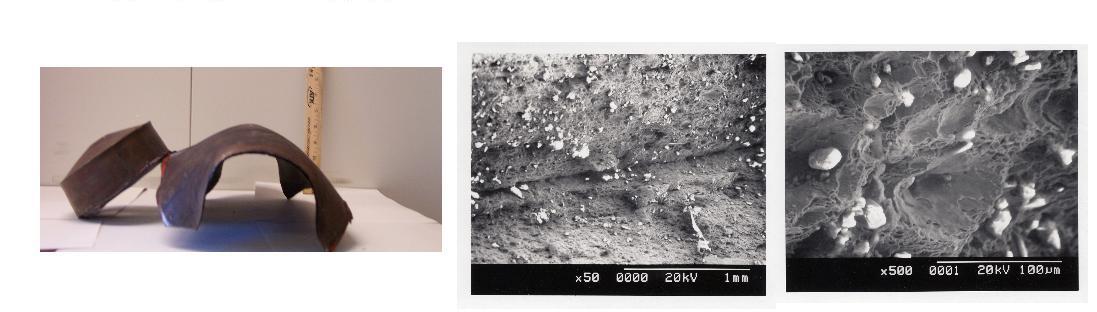}}\\
        (a) Ductile fracture/Void Growth and Coalescence\\\vspace{24pt}
    \scalebox{0.5}{\includegraphics{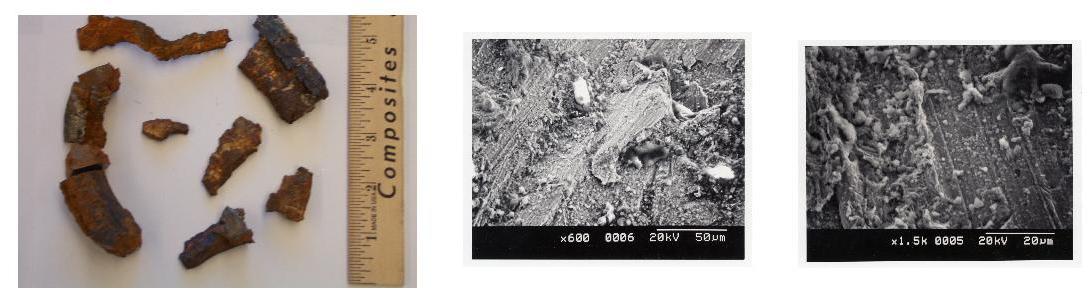}}\\
        (b) Fragmentation/Adiabatic Shear Bands
    \caption{Experimental tests of exploding cylinders.}
    \label{fig:container}
  \end{figure}

  Figure~\ref{fig:container} shows the recovered parts of AISI 1026 steel
  containers after two different tests.  The containers were initially 
  10 cm. in diameter and 0.6 cm thick.  In the first test, shown in 
  Figure~\ref{fig:container}(a), the container was heated over an open pool 
  fire and shows ductile failure.  In the second test, shown in 
  Figure~\ref{fig:container}(b), the container was heated by means of electrical
  tape and fragmented after the explosion.
  
  The dynamics of the solid materials - steel and PBX 9501 - is modeled 
  using the Lagrangian Material Point Method (MPM)~\cite{Sulsky94}.  Gases 
  are generated from solid PBX 9501 using a burn model~\cite{Long02}.  
  Gas-solid interaction is accomplished using an Implicit Continuous 
  Eulerian (ICE) multi-material hydrodynamic code~\cite{Guilkey04}.  A 
  single computational grid is used for all the materials.

  The constitutive response of PBX 9501 is modeled using 
  ViscoSCRAM~\cite{Bennet98}, which is a five element generalized Maxwell 
  model for the viscoelastic response coupled with statistical crack 
  mechanics.  Solid PBX 9501 is progressively converted into a gas with an 
  appropriate equation of state.  The temperature and pressure in the gas 
  increase rapidly as the reaction continues.  As a result, the steel container
  is pressurized, undergoes plastic deformation, and finally fragments.  The 
  entire process was simulated using the massively parallel, Common Component 
  Architecture ~\cite{Armstrong99} based, Uintah Computational 
  Framework (UCF)~\cite{Dav2000}.

  The main issues regarding the constitutive modeling of the steel container 
  are the selection of appropriate models for nonlinear elasticity, plasticity,
  damage, loss of material stability, and failure.  The numerical simulation 
  of the steel container involves the choice of appropriate algorithms for the
  integration of balance laws and constitutive equations, as well as the
  methodology for fracture simulation.  Models and simulation methods
  for the steel container are required to be temperature sensitive and
  valid for large distortions, large rotations, and  a range of strain 
  rates (quasistatic at the beginning of the simulation to approximately 
  $10^6$ s$^{-1}$ at fracture). 

  The approach chosen for the present work is to use hypoelastic-plastic
  constitutive models that assume an additive decomposition of the rate
  of deformation tensor into elastic and plastic parts.  Hypoelastic materials
  are known not to conserve energy in a loading-unloading cycle
  unless a very small time step is used.  However, the choice of this model 
  is justified under the assumption that elastic strains are expected to be 
  small for the problem under consideration and unlikely to affect the 
  computation significantly.  

  Two plasticity models for flow stress are considered 
  along with a two different yield conditions.  Explicit fracture simulation 
  is computationally expensive and prohibitive in the large simulations 
  under consideration.  The choice, therefore, has been to use damage models 
  and stability criteria for the prediction of failure (at material points) 
  and particle erosion for the simulation of fracture propagation.  

  The outline of the paper is as follows.  A brief description of the Material
  Point Method is given in Section~\ref{sec:mpm}.  The stress update algorithm
  and the how various plasticity models, yield conditions, equations of state
  etc. are used during the stress update are discussed in 
  Section~\ref{sec:algo}.  The models used for the simulations are discussed
  in Section~\ref{sec:models}.  The results of some simulations are
  presented in Section~\ref{sec:simul} and conclusions are presented
  in Section~\ref{sec:conclude}.

\section{The Material Point Method}\label{sec:mpm}

  The Material Point Method (MPM) ~\cite{Sulsky94} is a particle method for 
  structural mechanics simulations.  In this method, the state variables 
  of the material are described on Lagrangian particles or "material points".
  In addition, a regular, structured Eulerian grid is used as a computational 
  scratch pad to compute spatial gradients and to solve the governing
  conservation equations.  An explicit time-stepping version of the Material 
  Point Method has been used in the simulations presented in this paper.  
  The MPM algorithm is summarized below~\cite{Sulsky95}.  

  It is assumed that an particle state at the beginning of a time step 
  is known.  The mass ($m$), external force ($\Bf^{\Text}$), and 
  velocity ($\Bv$) of the particles are interpolated to the grid using 
  the relations
  \begin{equation}\label{eq:1}
    m_g = \sum_{p} S_{gp}~m_p ~,~~~~
    \Bv_g = (1/m_g)\sum_{p} S_{gp}~m_p~\Bv_p ~,~~~~
    \Bf^{\Text}_g = \sum_{p} S_{gp}~\Bf^{\Text}_p
  \end{equation}
  where the subscript ($g$) indicates a quantity at a grid node and a 
  subscript ($p$) indicates a quantity on a particle.  The symbol $\sum_p$
  indicates a summation over all particles.  The quantity ($S_{gp}$) is 
  the interpolation function of node ($g$) evaluated at the position of 
  particle ($p$).  Details of the interpolants used can be found 
  elsewhere~\cite{Bard04}.

  Next, the velocity gradient at each particle is computed using the 
  grid velocities using the relation
  \begin{equation} \label{eq:2}
    \Grad{\Bv_p} = \sum_g \BG_{gp} \Bv_g
  \end{equation}
  where $\BG_{gp}$ is the gradient of the shape function of node ($g$)
  evaluated at the position of particle ($p$).  The velocity gradient at 
  each particle is used to determine the Cauchy stress ($\Bsig_p$) at the
  particle using a stress update algorithm.

  The internal force at the grid nodes ($\Bf^{\Tint}_g$) is calculated 
  from the divergence of the stress using
  \begin{equation} \label{eq:3}
    \Bf^{\Tint}_g = \sum_p \BG_{gp}~\Bsig_p~V_p
  \end{equation}
  where $V_p$ is the particle volume.
  
  The equation for the conservation of linear momentum is next solved on
  the grid.  This equation can be cast in the form
  \begin{equation} \label{eq:4}
    \Bm_g ~ a_g = \Bf^{\Text}_g - \Bf^{\Tint}_g
  \end{equation}
  where $\Ba_g$ is the acceleration vector at grid node ($g$).  

  The velocity vector at node ($g$) is updated using an explicit (forward
  Euler) time integration, and the particle velocity and position are then
  updated using grid quantities.  The relevant equations are
  \begin{align} 
     \Bv_g(t+\Delta t) & = \Bv_g(t) + \Ba_g~\Delta t \label{eq:5} \\
     \Bv_p(t+\Delta t) & = \Bv_p(t) + \sum_g S_{gp}~\Ba_g~\Delta t ~;~~~~
     \Bx_p(t+\Delta t) = \Bx_p(t) + \sum_g S_{gp}~\Bv_g~\Delta t \label{eq:6}
  \end{align}

  The above sequence of steps is repeated for each time step.  The above 
  algorithm leads to particularly simple mechanisms for handling contact.
  Details of these contact algorithms can be found elsewhere~\cite{Bard01}.

\section{Plasticity and Failure Simulation}\label{sec:algo}

  A hypoelastic-plastic, semi-implicit approach~\cite{Zocher00} 
  has been used for the stress update in the simulations presented in this 
  paper.  An additive decomposition of the rate of deformation tensor into
  elastic and plastic parts has been assumed.  One advantage of this approach
  is that it can be used for both low and high strain rates.  Another advantage 
  is that many strain-rate and temperature-dependent plasticity and damage 
  models are based on the assumption of additive decomposition of strain rates,
  making their implementation straightforward.  

  The stress update is performed in a co-rotational frame which is equivalent 
  to using the Green-Naghdi objective stress rate.  An incremental update of 
  the rotation tensor is used instead of a direct polar decomposition of the 
  deformation gradient.  The accuracy of model is good if elastic strains are 
  small compared to plastic strains and the material is not unloaded.
  It is also assumed that the stress tensor can be divided into a volumetric 
  and a deviatoric component. The plasticity model is used to update only the 
  deviatoric component of stress assuming isochoric behavior.  The hydrostatic 
  component of stress is updated using a solid equation of state.

  Since the material in the container may unload locally after fracture, the 
  hypoelastic-plastic stress update may not work accurately under certain 
  circumstances.  An improvement would be to use a hyperelastic-plastic stress 
  update algorithm.  Also, the plasticity models are temperature dependent.
  Hence there is the issue of severe mesh dependence due to change of the
  governing equations from hyperbolic to elliptic in the softening regime
  ~\cite{Hill75,Bazant85,Tver90}.  Viscoplastic stress update models or 
  nonlocal/gradient plasticity models~\cite{Ramaswamy98,Hao00} can be used 
  to eliminate some of these effects and are currently under investigation. 
 
  A particle is tagged as "failed" when its temperature is greater than the
  melting point of the material at the applied pressure.  An additional
  condition for failure is when the porosity of a particle increases beyond a
  critical limit.  A final condition for failure is when a bifurcation 
  condition such as the Drucker stability postulate is satisfied.  Upon failure,
  a particle is either removed from the computation by setting the stress to
  zero or is converted into a material with a different velocity field 
  which interacts with the remaining particles via contact.  Either approach
  leads to the simulation of a newly created surface.

  In the parallel implementation of the stress update algorithm, sockets have 
  been added to allow for the incorporation of a variety of plasticity, damage, 
  yield, and bifurcation models without requiring any change in the stress 
  update code.  The algorithm is shown in Algorithm~\ref{algo1}.  The
  equation of state, plasticity model, yield condition, damage model, and
  the stability criterion are all polymorphic objects created using a 
  factory idiom in C++~\cite{Coplien92}.
  
  \begin{algorithm}[t]
    \caption{Stress Update Algorithm} \label{algo1}
    \KwData{{\bf Persistent}:Initial moduli, temperature, porosity, 
              scalar damage, equation of state, plasticity model, 
              yield condition, stability criterion, damage model\\
             {\bf Temporary}:Particle state at time $t$}
    \KwResult{Particle state at time $t+\Delta t$}

    \SetAlgoLined
    \For{all the patches in the domain}{
      Read the particle data and initialize updated data storage\;
      \For{all the particles in the patch} {
        Compute the velocity gradient, the rate of deformation tensor 
          and the spin tensor\;
        Compute the updated left stretch tensor, rotation tensor, and
          deformation gradient\;
        Rotate the input Cauchy stress and the rate of deformation tensor
          to the material configuration\;
        Compute the current shear modulus and melting temperature\;
        Compute the pressure using the equation of state, update the
        hydrostatic stress, and compute the trial deviatoric stress\;
        Compute the flow stress using the plasticity model\;
        Evaluate the yield function\;
        \eIf{particle is elastic}{
          Rotate the stress back to laboratory coordinates\;
          Update the particle state\;
        }{
          Find derivatives of the yield function\;
          Do radial return adjustment of deviatoric stress\;
          Compute updated porosity, scalar damage, and temperature 
           increase due to plastic work\;
          Compute elastic-plastic tangent modulus and evaluate stability
          condition\;
          Rotate the stress back to laboratory coordinates\;
          Update the particle state\;
          \If{Temperature $>$ Melt Temperature or Porosity $>$ Critical Porosity
               or Unstable}{
            Tag particle as failed\;
          }
        }
      }
    }
    Convert failed particles into a material with a different velocity field;
  \end{algorithm}
      
\section{Models}\label{sec:models}
  The stress in the solid is partitioned into a volumetric part and a deviatoric
  part.  Only the deviatoric part of stress is used in the plasticity 
  calculations assuming isoschoric plastic behavior.

  The hydrostatic pressure ($p$) is calculated either using the bulk modulus
  ($K$) and shear modulus ($\mu$) or from a temperature-corrected Mie-Gruneisen 
  equation of state of the form~\cite{Zocher00} 
  \begin{equation}
   p = \frac{\rho_0 C_0^2 \zeta
              \left[1 + \left(1-\frac{\Gamma_0}{2}\right)\zeta\right]}
             {\left[1 - (S_{\alpha} - 1) \zeta\right]^2 + \Gamma_0 C_p T}~,~~~~
   \zeta = (\rho/\rho_0 - 1)
  \end{equation}
  where $C_0$ is the bulk speed of sound, 
  $\rho_0$ is the initial density, $\rho$ is the current density, 
  $C_p$ is the specific heat at constant volume, $T$ is the temperature, 
  $\Gamma_0$ is the Gruneisen's gamma at reference state, and $S_{\alpha}$ 
  is the linear Hugoniot slope coefficient.

  Depending on the plasticity model being used, the pressure and
  temperature-dependent shear modulus ($\mu$) and the pressure-dependent 
  melt temperature ($T_m$) are calculated using the relations~\cite{Steinberg80}
  \begin{align}
    \mu & = \mu_0\left[1 + A\frac{p}{\eta^{1/3}} - B(T - 300)\right] \\
    T_m & = T_{m0} \exp\left[2a\left(1-\frac{1}{\eta}\right)\right]
              \eta^{2(\Gamma_0-a-1/3)}
  \end{align}
  where, $\mu_0$ is the shear modulus at the reference state($T$ = 300 K, 
  $p$ = 0, $\epsilon_p$ = 0), $\epsilon_p$ is the plastic strain.
  $\eta = \rho/\rho_0$ is the compression, $A = (1/\mu_0)(d\mu/dp)$, 
  $B = (1/\mu_0)(d\mu/dT)$, $T_{m0}$ is the melt temperature at 
  $\rho=\rho_0$, and $a$ is the coefficient of the first order volume 
  correction to Gruneisen's gamma.

  We have explored two temperature and strain rate dependent plasticity models -
  the Johnson-Cook plasticity model~\cite{Johnson83} and the Mechanical 
  Threshold Stress (MTS) plasticity model~\cite{Follans88,Goto00a}.  The
  flow stress ($\sigma_f$) from the Johnson-Cook model is given by
  \begin{equation}
    \sigma_f = [A + B (\epsilon_p)^n][1 + C \ln(\dot{\epsilon_p^*})]
    [1 - (T^*)^m]~;~~
    \dot{\epsilon_p^{*}} = \cfrac{\dot{\epsilon_p}}{\dot{\epsilon_{p0}}}~;~~
    T^* = \cfrac{(T-T_r)}{(T_m-T_r)}
  \end{equation}
  where $\dot{\epsilon_{p0}}$ is a user defined plastic strain rate, 
  A, B, C, n, m are material constants, $T_r$ is the room temperature, and
  $T_m$ is the melt temperature.  

  The flow stress for the MTS model is given by
  \begin{equation}
    \sigma_f = \sigma_a + \frac{\mu}{\mu_0} S_i \hat\sigma_i
    + \frac{\mu}{\mu_0} S_e \hat\sigma_e 
  \end{equation}
  where
  \begin{align*}
    \mu &= \mu_0 - \frac{D}{\exp\left(\frac{T_0}{T}\right) - 1} \\
    S_i &= \left[1 - \left(\frac{kT}{g_{0i}\mu b^3}
    \ln\frac{\dot\epsilon_{0i}}{\dot\epsilon}\right)^{1/qi}
    \right]^{1/pi} ~;~~
    S_e = \left[1 - \left(\frac{kT}{g_{0e}\mu b^3}
    \ln\frac{\dot\epsilon_{0e}}{\dot\epsilon}\right)^{1/qe}
    \right]^{1/pe}
  \end{align*}
  \begin{align*}
    \theta &= \theta_0 [ 1 - F(X)] + \theta_{IV} F(X) ~;~~ 
    \theta_0 = a_0 + a_1 \ln \dot\epsilon + a_2 \sqrt{\dot\epsilon} - a_3 T \\
    X & = \cfrac{\hat\sigma_e}{\hat\sigma_{es}} ~;~~ F(X) = \tanh(\alpha X) ~;~~
    \ln(\hat\sigma_{es}/\hat\sigma_{es0}) =
    \left(\frac{kT}{\mu b^3 g_{0es}}\right)
    \ln\left(\cfrac{\dot\epsilon}{\dot\epsilon_{es0}}\right)\\
    \hat\sigma_e^{(n+1)} & = \hat\sigma_e^{(n)}+\theta\Delta\epsilon
  \end{align*}
  and $\sigma_a$ is the athermal component of mechanical threshold stress,
  $\mu_0$ is the shear modulus at 0 K, $D, T_0$ are empirical constants, 
  $\hat\sigma_i$ represents the stress due to intrinsic barriers 
  to thermally activated dislocation motion and dislocation-dislocation 
  interactions, $\hat\sigma_e$ represents the stress due to 
  microstructural evolution with increasing deformation, 
  $k$ is the Boltzmann constant, $b$ is the length of the Burger's vector, 
  $g_{0[i,e]}$ are the normalized activation energies, 
  $\dot\epsilon_{0[i,e]}$ are constant strain rates,
  $q_{[i,e]}, p_{[i,e]}$ are constants, $\theta_0$ is the hardening due to 
  dislocation accumulation, $a_0, a_1, a_2, a_3, \theta_{IV}, \alpha$ are 
  constants,
  $\hat\sigma_{es}$ is the stress at zero strain hardening rate, 
  $\hat\sigma_{es0}$ is the saturation threshold stress for deformation at 0 K,
  $g_{0es}$ is a constant, and $\dot\epsilon_{es0}$ is the maximum strain rate.

  We have decided to focus on ductile failure of the steel container.
  Accordingly, two yield criteria have been explored - the von Mises condition
  and the Gurson-Tvergaard-Needleman (GTN) yield 
  condition~\cite{Gurson77,Tver84} which depends on porosity.  An associated 
  flow rule is used to determine the plastic rate parameter in either case.
  The von Mises yield condition is given by
  \begin{equation}
    \Phi = \left(\frac{\sigma_{eq}}{\sigma_f}\right)^2 - 1 = 0 ~;~~~
    \sigma_{eq} = \sqrt{\frac{3}{2}\sigma^{d}:\sigma^{d}}
  \end{equation}
  where $\sigma_{eq}$ is the von Mises equivalent stress, 
  $\sigma^{d}$ is the deviatoric part of the Cauchy stress, and
  $\sigma^{f}$ is the flow stress.
  The GTN yield condition can be written as
  \begin{equation}
    \Phi = \left(\frac{\sigma_{eq}}{\sigma_f}\right)^2 +
    2 q_1 f_* \cosh \left(q_2 \frac{Tr(\sigma)}{2\sigma_f}\right) -
    (1+q_3 f_*^2) = 0
  \end{equation}
  where $q_1,q_2,q_3$ are material constants and $f_*$ is the porosity 
  (damage) function given by
  \begin{equation}
    f* = 
    \begin{cases}
      f & \text{for}~~ f \le f_c,\\ 
      f_c + k (f - f_c) & \text{for}~~ f > f_c 
    \end{cases}
  \end{equation}
  where $k$ is a constant and $f$ is the porosity (void volume fraction).  The 
  flow stress in the matrix material is computed using either of the two 
  plasticity models discussed earlier.  Note that the flow stress in the matrix 
  material also remains on the undamaged matrix yield surface and uses an 
  associated flow rule.

  The evolution of porosity is calculated as the sum of the rate of growth 
  and the rate of nucleation~\cite{Ramaswamy98a}.  The rate of growth of
  porosity and the void nucleation rate are given by the following equations
  ~\cite{Chu80}
  \begin{align}
    \dot{f} &= \dot{f}_{\text{nucl}} + \dot{f}_{\text{grow}} \\
    \dot{f}_{\text{grow}} & = (1-f) \text{Tr}(\BD_p) \\
    \dot{f}_{\text{nucl}} & = \cfrac{f_n}{(s_n \sqrt{2\pi})}
            \exp\left[-\Half \cfrac{(\epsilon_p - \epsilon_n)^2}{s_n^2}\right]
            \dot{\epsilon}_p
  \end{align}
  where $\BD_p$ is the rate of plastic deformation tensor, $f_n$ is the volume 
  fraction of void nucleating particles , $\epsilon_n$ is the mean of the 
  distribution of nucleation strains, and $s_n$ is the standard 
  deviation of the distribution.

  Part of the plastic work done is converted into heat and used to update the 
  temperature of a particle.  The increase in temperature ($\Delta T$) due to 
  an increment in plastic strain ($\Delta\epsilon_p$) is given by the equation
  ~\cite{Borvik01}
  \begin{equation}
    \Delta T = \cfrac{\chi\sigma_f}{\rho C_p} \Delta \epsilon_p
  \end{equation}
  where $\chi$ is the Taylor-Quinney coefficient, and $C_p$ is the specific
  heat.  A special equation for the dependence of $C_p$ upon temperature is
  also used for steel~\cite{Goto00}.
  \begin{equation}
    C_p = 10^3(0.09278 + 7.454\times 10^{-4} T + 12404.0/T^2)
  \end{equation}

  Under normal conditions, the heat generated at a material point is conducted 
  away at the end of a time step using the heat equation.  If special adiabatic 
  conditions apply (such as in impact problems), the heat is accumulated at a 
  material point and is not conducted to the surrounding particles.  This 
  localized heating can be used to simulate adiabatic shear band formation.

  After the stress state has been determined on the basis of the yield condition
  and the associated flow rule, a scalar damage state in each material point can
  be calculated using either of two damage models - the Johnson-Cook model
  ~\cite{Johnson85} or the Hancock-MacKenzie model~\cite{Hancock76}.  While 
  the Johnson-Cook model has an explicit dependence on temperature, 
  the Hancock-McKenzie model depends on the temperature implicitly, via the 
  stress state.  Both models depend on the strain rate to determine the 
  value of the scalar damage parameter.

  The damage evolution rule for the Johnson-Cook damage model can be written as
  \begin{equation}
    \dot{D} = \cfrac{\dot{\epsilon_p}}{\epsilon_p^f} ~;~~
    \epsilon_p^f = 
      \left[D_1 + D_2 \exp \left(\cfrac{D_3}{3} \sigma^*\right)\right]
      \left[1+ D_4 \ln(\dot{\epsilon_p}^*)\right]
      \left[1+D_5 T^*\right]~;~~
    \sigma^*= \cfrac{\text{Tr}(\Bsig)}{\sigma_{eq}}~;~~
  \end{equation}
  where $D$ is the damage variable which has a value of 0 for virgin material
  and a value of 1 at fracture, $\epsilon_p^f$ is the fracture strain, 
  $D_1, D_2, D_3, D_4, D_5$ are constants, $\Bsig$ is the Cauchy stress, and
  $T^*$ is the scaled temperature as in the Johnson-Cook plasticity model.

  The Hancock-MacKenzie damage evolution rule can be written as
  \begin{equation}
    \dot{D} = \cfrac{\dot{\epsilon_p}}{\epsilon_p^f} ~;~~
    \epsilon_p^f = \frac{1.65}{\exp(1.5\sigma^*)}
  \end{equation}

  The determination of whether a particle has failed can be made on the 
  basis of either or all of the following conditions:
  \begin{itemize}
    \item The particle temperature exceeds the melting temperature.
    \item The TEPLA-F fracture condition~\cite{Johnson88} is satisfied.
       This condition can be written as
       \begin{equation}
         (f/f_c)^2 + (\epsilon_p/\epsilon_p^f)^2 = 1
       \end{equation}
       where $f$ is the current porosity, $f_c$ is the maximum 
       allowable porosity, $\epsilon_p$ is the current plastic strain, and
       $\epsilon_p^f$ is the plastic strain at fracture.
    \item An alternative to ad-hoc damage criteria is to use the concept of 
       bifurcation to determine whether a particle has failed or not.  Two
       stability criteria have been explored in this paper - the Drucker
       stability postulate~\cite{Drucker59} and the loss of hyperbolicity
       criterion (using the determinant of the acoustic tensor)
       \cite{Rudnicki75,Perzyna98}.  
  \end{itemize}

  The simplest criterion that can be used is the Drucker stability postulate 
  \cite{Drucker59} which states that time rate of change of the rate of 
  work done by a material cannot be negative.  Therefore, the material is 
  assumed to become unstable (and a particle fails) when
  \begin{equation}
    \dot\Bsig:\BD^p \le 0
  \end{equation}

  Another stability criterion that is less restrictive is the acoustic
  tensor criterion which states that the material loses stability if the 
  determinant of the acoustic tensor changes sign~\cite{Rudnicki75,Perzyna98}.  
  Determination of the acoustic tensor requires a search for a normal vector 
  around the material point and is therefore computationally expensive.  A 
  simplification of this criterion is a check which assumes that the direction 
  of instability lies in the plane of the maximum and minimum principal 
  stress~\cite{Becker02}.  In this approach, we assume that the strain is 
  localized in a band with normal $\Bn$, and the magnitude of the velocity 
  difference across the band is $\Bg$.  Then the bifurcation condition 
  leads to the relation 
  \begin{equation} 
    R_{ij} g_{j} = 0 ~;~~~
    R_{ij} = M_{ikjl} n_k n_l + M_{ilkj} n_k n_l - \sigma_{ik} n_j n_k
  \end{equation} 
  where $M_{ijkl}$ are the components of the co-rotational tangent
  modulus tensor and $\sigma_{ij}$ are the components of the co-rotational 
  stress tensor.  If $\det(R_{ij}) \le 0 $, then $g_j$ can be arbitrary and 
  there is a possibility of strain localization.  If this condition for 
  loss of hyperbolicity is met,  then a particle deforms in an unstable 
  manner and failure can be assumed to have occurred at that particle.  

\section{Simulations}\label{sec:simul}
  
\begin{figure}[t]
    \centering
    \scalebox{0.4}{\rotatebox{-90}{\includegraphics{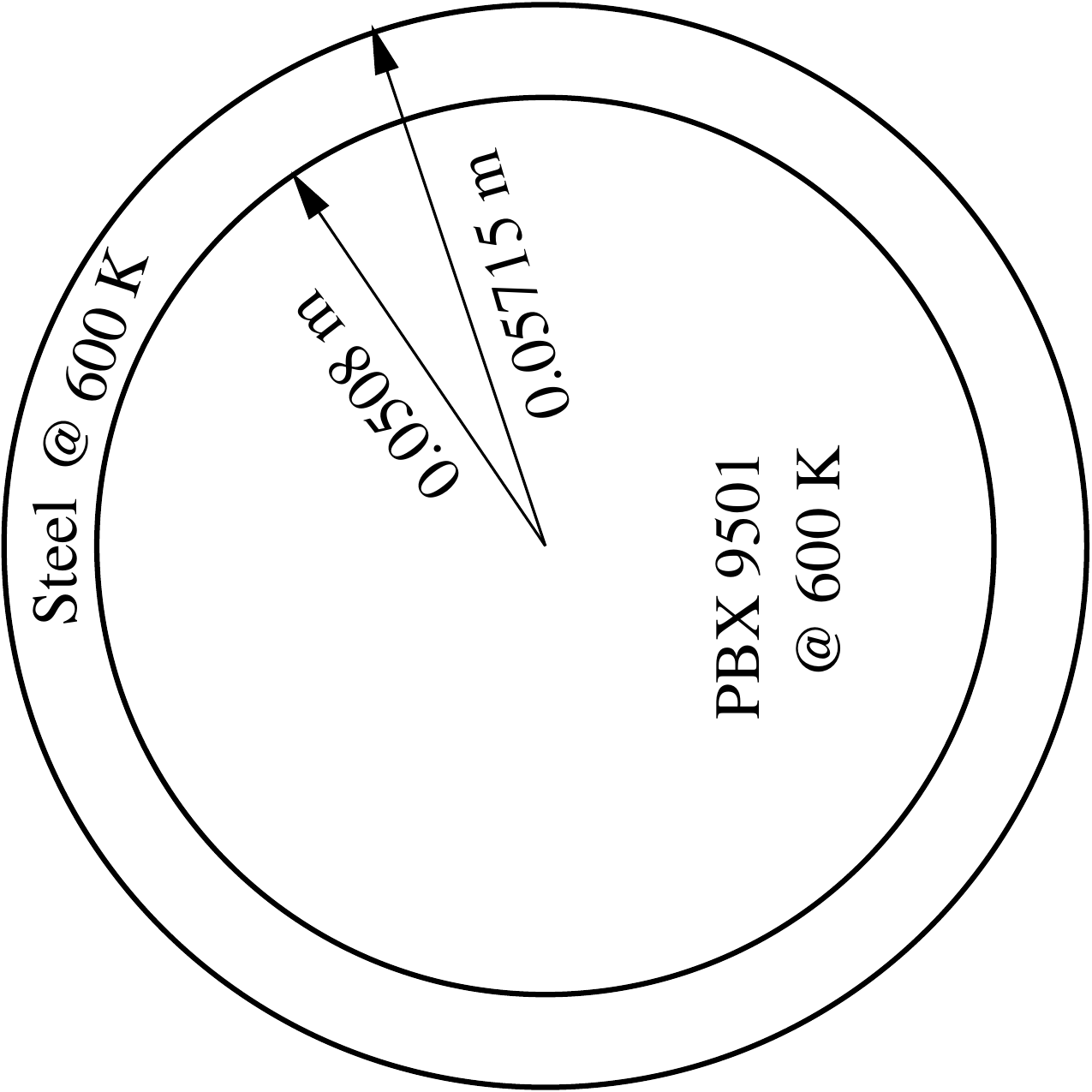}}}\\
    (a) Geometry 
    \vspace{12pt}
    \scalebox{0.4}{\includegraphics{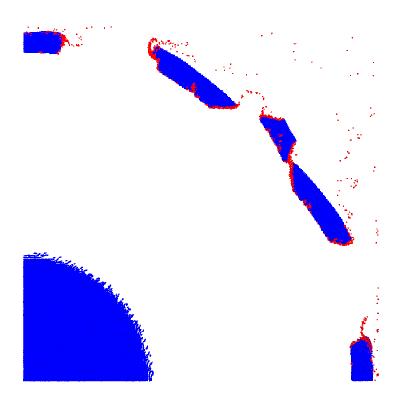} \hspace{1in}
                   \hspace{12pt}
                   \includegraphics{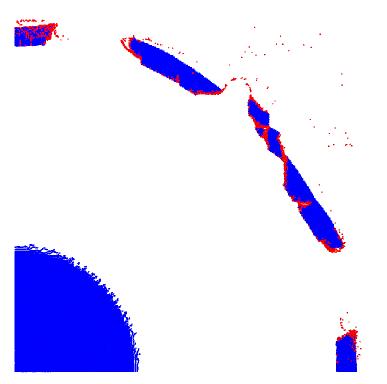}}\\
    (b) 4340 Steel. \hspace{1in} (c) HY 100 Steel.
    \caption{Simulations of fragmenting cylinders - two-dimensional view.}
    \label{fig:fragments}
  \end{figure}

  The first set of simulations was performed using the geometry shown in 
  Figure~\ref{fig:fragments}(a).  A steel cylinder was used to confine the 
  PBX 9501 material and the simulation was started with both materials at 
  a temperature of 600 K.  At this temperature, PBX 9501 reacts and forms 
  gases which expand the cylinder.
  A quarter of the cylinder was modeled using a $160\times 160\times 1$ grid
  with 8 particles per grid cell.  The shapes of the cylinder after failure
  for two different materials are shown in Figure~\ref{fig:fragments}.

  The simulation shown in Figure~\ref{fig:fragments}(b) was performed using 
  material data for 4340 steel, a Mie-Gr\"{u}neisen equation of state,
  the Johnson-Cook flow stress model, the Gurson yield condition, the 
  Johnson-Cook damage model, and checks of both the Drucker stability 
  postulate and the loss of hyperbolicity condition.  The simulation of
  a HY 100 steel container shown in Figure~\ref{fig:fragments}(c) was 
  performed using a Mie-Gr\"{u}neisen equation of state, the MTS flow stress
  model, the Gurson yield condition, the Hancock-MacKenzie damage model,
  and the same stability checks as the 4340 steel.  The material properties
  and the parameters used in the models are shown in Table~\ref{tab:matprop}.
  The materials are given an initial mean porosity of 0.005 using a Gaussian
  distribution with a standard deviation of 0.001 and a mean scalar damage
  value of 0.01 with a standard deviation of 0.005.
  \begin{table}[t]
    \caption{Material Properties and Parameters for Steels.}
    \label{tab:matprop}
    \centering
    \small
    \renewcommand{\arraystretch}{1.25}
    \begin{tabular}{cccccccccc}
    \hline\hline
    \multicolumn{10}{c}{4340 Steel properties and Johnson-Cook parameters}\\
    \hline
    $\rho$    & $C_p$            & $T_m$  & $K$   & $\mu$ & $\chi$ \\
    (kg/m$^3$)& (MPa m$^3$/kg K) & (K)    & (GPa) & (GPa) & \\
    \hline
    7830.0    & 477.0            & 1793.0 & 173.3 & 80.0 & 0.9\\
    \hline
    $A$   & $B$   & $C$   & $n$  & $m$  & $D_1$ & $D_2$ & $D_3$ & $D_4$ &$D_5$\\
    (MPa) & (MPa) &       &      &      &       &       &       &       & \\
    \hline
    792.0 & 510.0 & 0.014 & 0.26 & 1.03 & 0.05  & 3.44  & -2.12 & 0.002 & 0.61\\
    \hline
    \hline
    \multicolumn{10}{c}{HY100 Steel properties and MTS parameters}\\
    \hline
    $\rho$    & $C_p$            & $T_m$  & $K$   & $\mu$ & $\chi$\\
    (kg/m$^3$)& (MPa m$^3$/kg K) & (K)    & (GPa) & (GPa) & \\
    \hline
    7860.0    & 477.0            & 2000.0 & 150.0 & 69.0 & 0.9\\
    \hline
    $\Sa$ & $\mu_0$ & $D$   & $T_0$ & $k/b^3$  & $\Gi$ & $\Ge$ & $\Ei$       & 
      $\Ee$    \\
    (MPa) & (GPa)   & (GPa) & (K)   & (x$10^6$)&       &       & (x$10^{13}$)&
      (x$10^7$)\\
    \hline
    40.0  & 71.46   & 2.9   & 204   & 0.905    & 1.161 & 1.6   & 1.0         &
      1.0      \\
    \hline
    $\Pci$ & $\Qci$ & $\Pce$ & $\Qce$ & $\Shi$ & $a_0$ & $a_1$ & $a_2$ & $a_3$\\
           &        &        &        & (MPa) & (x$10^9$)&   &  & (x$10^6$)\\
    \hline
    0.5 & 1.5 & 0.67 & 1.0 & 1341 & 6 & 0 & 0 & 2.0758 \\
    \hline
    $\TIV$    & $\alpha$ & $\Ees$    & $\Ges$ & $\Ses$ \\
    (x$10^6$) &          & (x$10^7$) &        & (MPa) \\
    \hline
    200.0     & 3 & 1.0 & 0.112 & 822.0 \\
    \hline
    \hline
    \multicolumn{10}{c}{Mie-Gruneisen equation of state parameters}\\
    \hline
    $C_0$ & $\Gamma_0$ & $S_{\alpha}$ \\
    (m/s) &            &              \\
    \hline
    3574  & 1.69 & 1.92 \\
    \hline
    \hline
    \multicolumn{10}{c}{GTN yield condition and porosity evolution parameters}\\
    \hline
    $q_1$ & $q_2$ & $q_3$ & $k$ & $f_c$ & $f_n$ & $s_n$ & $\En$ \\
    \hline
    \hline
     1.5 & 1.0 & 2.25 & 4.0 & 0.05 & 0.1 & 0.3 & 0.1\\
    \hline
    \hline
    \end{tabular}
    \normalsize
  \end{table}

  The expected number of fragments ($N$) along the circumference of the 
  exploding cylinder can be approximated using the following analytical 
  result~\cite{Grady92}
  \begin{equation}
   N = 2~\pi~\left(\frac{\rho~R_0~V^2}{24~\Gamma}\right)^{1/3}
  \end{equation}
  where $\rho$ is the density, $R_0$ is the initial cylinder radius,
  $V$ is the expansion velocity at the radius of fracture, and
  $\Gamma$ is the fragmentation energy.

  The fragmentation energy in tension ($\Gamma_T$) and in shear ($\Gamma_S$)
  are given by
  \begin{equation}
  \Gamma_T = \frac{K_c^2}{2~E} ~;~~~
  \Gamma_S = \frac{\rho~C_p}{\alpha}
           \left(
            \frac{9~\rho^3~C_p^2~\chi^3}{Y^3~\alpha^2\dot{\gamma}}
           \right)^{1/4}
  \end{equation}
  where $K_c$ is the fracture toughness, $E$ is the Young's modulus,
  $\rho$ is the density, $C_p$ is the specific heat at constant pressure,
  $\alpha$ is the thermal softening coefficient, $\chi$ is the thermal diffusion
  coefficient, $Y$ is the yield strength in simple tension, and $\dot{\gamma}$
  is the shear strain rate.

  For the expanding 4340 steel cylinder of that we have simulated, the relevant 
  quantities are $\rho$ = 7830 kg/m$^3$, $E$ = 208 GPa, $K_c$ = 80 MN/m$^{2/3}$,
  $Y$ = 792 MPa, $C_p$ = 477 J/kg~K, $\chi$ = 1.5$\times 10^{-5}$ m$^2$/s, 
  $\alpha$ = 7.5$\times 10^{-4}$ /K, $R_0$ = 0.054 m, $V$ = 300 m/s,
  $\dot{\gamma}$ = 1000 /s, $\Gamma_T$ = 1.5 $\times 10^4$ J/m$^2$, and
  $\Gamma_S$ = 5.2 $\times 10^4$ J/m$^2$.  Accordingly, the expected number of 
  fragments (for the whole cylinder) are $N~(\text{tension})$ = 29 and
  $N~(\text{shear})$ = 20.  For a quarter of the cylinder, the number of 
  fragments is expected to be
  between 8 and 5.  We get approximately 6 to 7 fragments in our simulations,
  which implies that our results are qualitatively acceptable.  Both the 
  steels show similar fragmentation though the exact shape of the fragments
  differs slightly.  For this reason, the three-dimensional simulations were
  performed using 4340 steel and the associated models discussed above.

\begin{figure}
    \scalebox{1.0}{\includegraphics{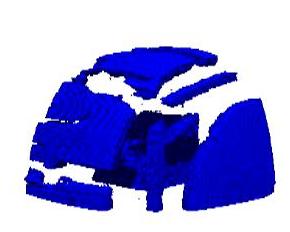} 
                   \hspace{12pt}
                   \includegraphics{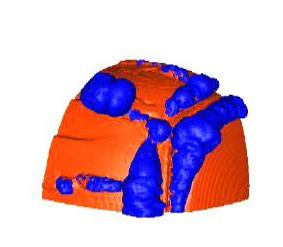}}\\
    (a) Fragments of the container. \hspace{1in} 
    (b) Gases escaping from the container.
    \caption{Simulations of fragmenting cylinders - three-dimensional view.}
    \label{fig:frag3D}
  \end{figure}

  Figure~\ref{fig:frag3D} shows the fragmentation obtained from 
  three-dimensional simulations of a cylinder with end-caps.  A quarter
  of the geometry is modeled, assuming symmetry.  The cylinder is made of
  4340 steel and contains PBX 9501.  The simulation is started with both
  materials at a temperature of 600 K.  A hypoelastic constitutive model
  is used to determine the volumetric response of the material.  The 
  Johnson-Cook plasticity model is used to calculate the flow stress.
  The von Mises yield condition is used to determine the boundary of the 
  elastic and plastic domains.  A Johnson-Cook damage model is used to 
  compute a scalar damage parameter.  A uniform initial porosity is assigned
  to all steel particles and evolved according to the models discussed in
  the previous section.  A particle is deemed to have failed when the 
  modified TEPLA-F condition is satisfied, the temperature is more than the
  melting temperature, or the Drucker stability postulate/loss of hyperbolicity
  condition is satisfied.  Upon failure, the particle stress is set to zero.

  The simulations capture some of the qualitative features observed in the
  experiments of steel cylinders heated using heat tapes.  Some high 
  particle velocities are observed upon failure.  Simulations have shown
  that these velocities are due to some increase in the total energy due to
  the setting of the particle stress to zero.  Computations where failed
  particles are converted into a material with a different velocity field
  are currently under way along with other validation efforts to quantify 
  the error in the calculations.
  
\begin{figure}
    \scalebox{0.38}{\includegraphics{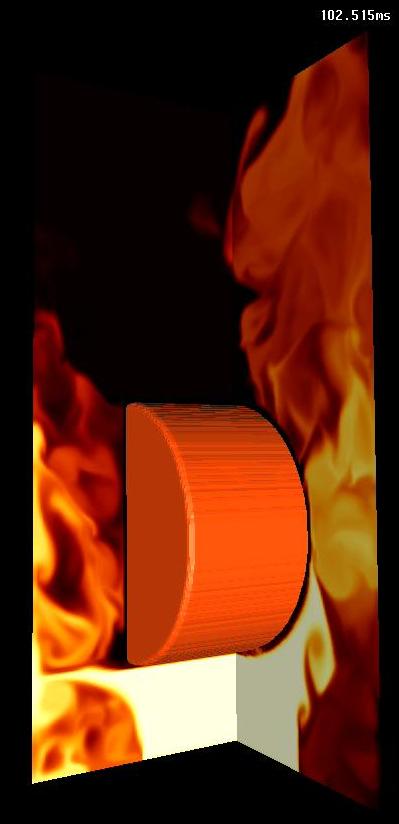} \hspace{12pt}
                   \includegraphics{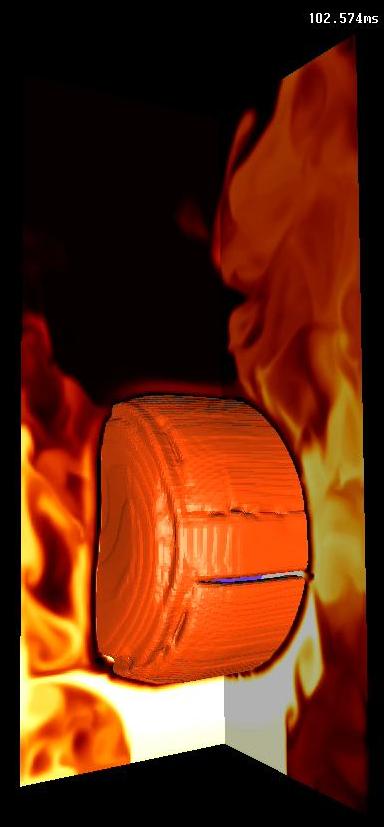} \hspace{12pt}
                   \includegraphics{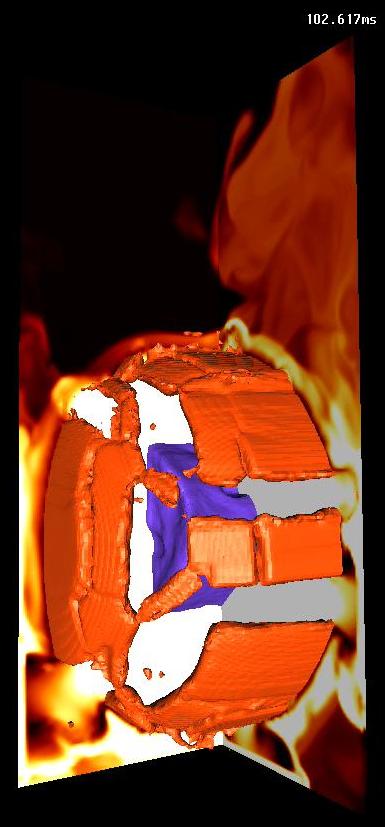} \hspace{12pt}
                   \includegraphics{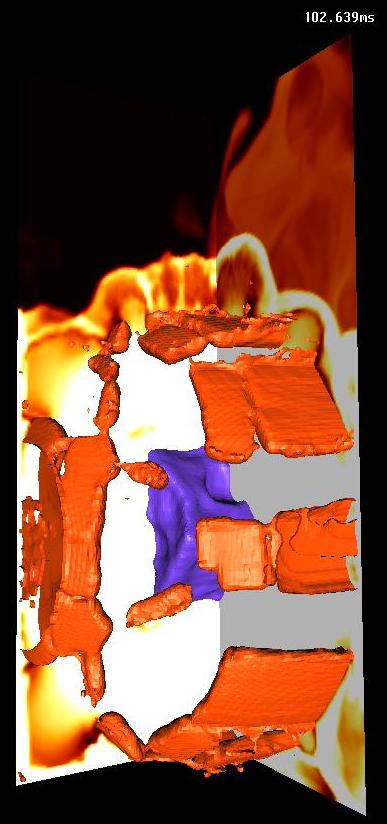}}\\
    \caption{Simulations of fragmentation of a cylinder heated by a fire.}
    \label{fig:poolFire}
  \end{figure}

  Simulations have also been performed on containers heated by a pool fire.
  Four snapshots of one such simulation for 4340 steel using the Johnson-Cook
  plasticity and damage models, a Mie-Gruneisen equation of state, the 
  von Mises yield condition and uniform initial porosity are shown
  in Figure~\ref{fig:poolFire}.
  The fire is simulated by a hot jet of air and interacts with the container
  via heat conduction and momentum transport.  After some time, the contents
  of the container reach ignition temperature and reaction proceeds rapidly.
  Axial cracks open up in the container that are qualitatively similar to 
  those observed in experiments.  These cracks join to form fragments which
  interact with the fire.  More details of such simulations can be found
  elsewhere~\cite{Guilkey04}.  Validation of the pool fire and container
  interaction scenario is currently being performed in collaboration with 
  researchers at the Lawrence Livermore National Laboratory.

  \section{Summary and Conclusions}\label{sec:conclude}
  A computational scheme for the simulation of the fragmentation of 
  cylinders due to interaction with gases from a reacting high energy 
  material has been presented.  The scheme allows for the incorporation 
  of various plasticity models, yield conditions, damage models, equations 
  of state, and stability checks within the same stress update code.
  A number of such models have been listed and the corresponding material
  properties and parameters for two steels have been collected from various 
  sources and presented in a compact form.  

  Simulations of exploding cylinders in two-dimensions have been compared 
  with analytical solutions for the expected number of fragments and found 
  to provide qualitative agreement.  Three-dimensional simulations also
  show qualitative agreement with experiments in the directions of the
  dominant cracks.  Snapshots from the simulation of a fully coupled
  container-fire simulation have also shown qualitative agreement with 
  experiments.  

  Two issues that have been identified as important for the simulation are
  the conservation of energy and mesh dependence of the results.  Validation
  simulations that are currently underway have shown that energy is better
  conserved when particles are converted into a material with a different 
  velocity field after failure (rather than setting the stress to zero
  upon failure).  Results of these tests will be presented in future work.
  In the absence of a limiting length scale in the computation,  strongly 
  mesh dependent behavior can be expected in the softening regime of the 
  stress-strain relationship.  This mesh dependence occurs when we use 
  temperature dependent elastic/plastic constitutive equations and when 
  we degrade the yield strength of the porous material using porosity. 
  One way of minimizing mesh dependence is to use a rate-dependent stress 
  update algorithm.  However, such an approach is not sufficient to remove 
  the effects of all the possible causes of mesh dependence.  Validation
  experiments are currently underway to determine the extent of mesh
  dependence and nonlocal/gradient plasticity approaches that can be 
  formulated for the material point method.  Overall, the material point
  method appears to be a promising approach for simulating high rate, coupled
  fluid-structure interaction problems and fragmentation.

\section{Acknowledgments}
  This work was sponsored by the Department of Energy Accelerated 
  Supercomputing Initiative (DOE-ASCI), Lawrence Livermore National
  Laboratory and the Center for the Simulation of Accidental Fires and
  Explosions (C-SAFE), University of Utah.  The author would like to 
  acknowledge Dr. Steve Parker and his team in the School of 
  Computing, University of Utah, for providing the infrastructure for
  parallel computing and visualization.  Thanks also go to Drs. James
  Guilkey and Todd Harman from the Department of Mechanical Engineering,
  University of Utah, for providing the fluid-structure interaction 
  code and performing the large container and fluid-structure interaction
  simulations on supercomputers at the Los Alamos National Laboratory. 
\newpage
%
%
\appendix
%
%
\bibliography{mybiblio}
%
%
%
\end{document}